\documentclass[twoside]{article}

\usepackage[T1]{fontenc}
\usepackage{tgtermes}

\usepackage{amssymb}
\newcommand{\mathbold}[1]{\ensuremath{\boldsymbol{\mathbf{#1}}}}

\usepackage{scalefnt,letltxmacro}
\LetLtxMacro{\oldtextsc}{\textsc}
\renewcommand{\textsc}[1]{\oldtextsc{\scalefont{1.10}#1}}
\usepackage[ttdefault=true]{AnonymousPro}

\usepackage{amsmath, amsthm, amsfonts}
\usepackage{mathtools}

\usepackage[
  paper  = letterpaper,
  left   = 1.50in,
  right  = 1.50in,
  top    = 1.0in,
  bottom = 1.0in,
  ]{geometry}
\usepackage[english]{babel}
\usepackage[parfill]{parskip}
\usepackage{afterpage}
\usepackage{enumitem}
\usepackage{framed}
\usepackage{xspace}

\usepackage{lineno}

\usepackage{ragged2e}

\newcounter{parcount}

\DeclareRobustCommand{\parhead}[1]{\textbf{#1}~}

\usepackage[usenames,dvipsnames]{xcolor}
\definecolor{shadecolor}{gray}{0.9}

\newcommand{\red}[1]{\textcolor{BrickRed}{#1}}

\usepackage{graphicx}
\usepackage[labelfont=bf]{caption}
\usepackage[format=hang]{subcaption}

\usepackage{booktabs, array}

\usepackage[algoruled]{algorithm2e}
\usepackage{listings}
\usepackage{fancyvrb}
\fvset{fontsize=\small}

\usepackage{natbib}
\usepackage[colorlinks,linktoc=all]{hyperref}
\hypersetup{citecolor=MidnightBlue}
\hypersetup{linkcolor=MidnightBlue}
\hypersetup{urlcolor=MidnightBlue}

\newcommand{\myeqp}[1]{\hyperref[eq:#1]{Eq.\ref*{eq:#1}}}
\newcommand{\mysub}[1]{\hyperref[sub:#1]{Section~\ref*{sub:#1}}}
\newcommand{\mysec}[1]{\hyperref[sec:#1]{Section~\ref*{sec:#1}}}
\newcommand{\mytable}[1]{\hyperref[table:#1]{Table~\ref*{table:#1}}}
\newcommand{\myfig}[1]{\hyperref[fig:#1]{Figure~\ref*{fig:#1}}}
\newcommand{\myappendix}[1]{\hyperref[appendix:#1]{Appendix~\ref*{appendix:#1}}}
\newcommand{\myalg}[1]{\hyperref[alg:#1]{Algorithm~\ref*{alg:#1}}}
\newcommand{\mytheorem}[1]{\hyperref[theorem:#1]{Theorem~\ref*{theorem:#1}}}
\newcommand{\myfootnote}[1]{\hyperref[footnote:#1]{Footnote~\ref*{footnote:#1}}}

\usepackage
[acronym,smallcaps,nowarn,section,nonumberlist]{glossaries}
\glsdisablehyper{}

\lstdefinestyle{mystyle}{
    commentstyle=\color{OliveGreen},
    numberstyle=\tiny\color{black!60},
    stringstyle=\color{BrickRed},
    basicstyle=\ttfamily\scriptsize,
    breakatwhitespace=false,
    breaklines=true,
    captionpos=b,
    keepspaces=true,
    numbers=none,
    numbersep=5pt,
    showspaces=false,
    showstringspaces=false,
    showtabs=false,
    tabsize=2
}
\RequirePackage[bf,small]{titlesec}

\newcommand{\g}{\,|\,}

\usepackage{bbm}
\newcommand{\ind}{\mathbbm{1}}

\renewcommand{\d}[1]{\ensuremath{\operatorname{d}\!{#1}}}

\newcommand\indep{\protect\mathpalette{\protect\independenT}{\perp}}
\def\independenT#1#2{\mathrel{\rlap{$#1#2$}\mkern2mu{#1#2}}}
\newcommand{\E}[1]{\mathbb{E}\left[ #1 \right]}
\newcommand{\EE}[2]{\mathbb{E}_{#1}\left[ #2 \right]}
\newcommand{\cD}{\mathcal{D}}
\newcommand{\rep}{\textrm{rep}}

\newcommand{\mba}{\mathbold{a}}

\newcommand{\mbx}{\mathbold{x}}
\newcommand{\mby}{\mathbold{y}}

\newcommand{\mbA}{\mathbold{A}}

\newcommand{\mbY}{\mathbold{Y}}

\newcommand{\mbphi}{\mathbold{\phi}}

\newcommand{\mbtheta}{\mathbold{\theta}}

\newacronym{VI}{vi}{variational inference}
\newacronym{KL}{kl}{Kullback-Leibler}
\newacronym{ELBO}{elbo}{\emph{evidence lower bound}}
\newacronym{MCMC}{mcmc}{Markov chain Monte Carlo}
\newacronym{ppc}{ppc}{posterior predictive check}
\newacronym{ate}{ate}{average treatment effect}

\usepackage{tikz}

\usetikzlibrary{bayesnet}

\pgfdeclarelayer{edgelayer}
\pgfdeclarelayer{nodelayer}
\pgfsetlayers{edgelayer,nodelayer,main}

\definecolor{hexcolor0xbfbfbf}{rgb}{0.749,0.749,0.749}

\tikzset{>=latex}
\tikzstyle{none}   = [inner sep=0pt]
\tikzstyle{line}   = [ -, thick, shorten <=1pt, shorten >=1pt ]
\tikzstyle{arrow}  = [ ->, thick, shorten <=1pt, shorten >=1pt ]
\tikzstyle{ardash} = [ dashed, ->, thick, shorten <=1pt, shorten >=1pt ]

\tikzstyle{empty}=[circle,opacity=0.0,text opacity=1.0,inner sep=0pt]
\tikzstyle{box}=[rectangle,fill=White,draw=Black]
\tikzstyle{filled}=[circle,thick,fill=hexcolor0xbfbfbf,draw=Black]
\tikzstyle{hollow}=[circle,thick,fill=White,draw=Black]
\tikzstyle{param}=[rectangle,fill=Black,draw=Black,inner sep=0pt,minimum width=4pt,minimum height=4pt]
\tikzstyle{paramhollow}=[rectangle,thick,fill=White,draw=Black,inner sep=0pt,minimum
width=4pt,minimum height=4pt]

\usepackage{pgfplots}                               
\pgfplotsset{compat=newest}
\pgfplotsset{plot coordinates/math parser=false}
\newlength\figureheight
\newlength\figurewidth
\newlength\figureheightsmall
\newlength\figurewidthsmall
\definecolor{POSTcolor}{rgb}{0.48, 0.20, 0.58} 
\definecolor{Qcolor}{rgb}{0.00, 0.53, 0.22} 

\usepackage[accepted]{aistats2017}

\begin{document}

\twocolumn[

\aistatstitle{Model Criticism for Bayesian Causal Inference}

\aistatsauthor{%
Dustin Tran \And Francisco J.R. Ruiz \And Susan Athey \And David M.
Blei
}

\aistatsaddress{%
Columbia University \And
Columbia University \And
Stanford University \And
Columbia University
}
]

\begin{abstract}
  The goal of causal inference is to understand the outcome of
  alternative courses of action. However, all causal inference
  requires assumptions.
  Such assumptions can be more influential than in typical
  tasks for probabilistic modeling, and testing those assumptions is important
  to assess the validity of causal inference. We develop model criticism
  for Bayesian causal inference, building on the idea of posterior
  predictive checks to assess model fit. Our approach involves
  decomposing the problem, separately criticizing the model of
  treatment assignments and the model of outcomes.
  Conditioned on the assumption of unconfoundedness---that the
  treatments are assigned independently of the potential outcomes---we
  show how to check any additional modeling assumption.
  Our approach provides a foundation for
  diagnosing model-based causal inferences.
\end{abstract}

\section{Introduction}

Consider the problem of understanding the
``treatment effect'' of an intervention, such as giving a drug to
patients with a specific disease.  In the language
of~\citet{neyman1923on,rubin1974estimating}, each individual has a
potential outcome when given the drug and a potential outcome when not
given the drug.  One measurement of the causal effect is the average
difference (over individuals) between those potential outcomes.  In
the language of graphical models, this is
framed as evaluating the impact of an intervention on random variables
in a probabilistic graph~\citep{pearl2000causality}.  The fundamental problem of causal
inference is that we do not observe both potential outcomes for any
individual at the same time~\citep{holland1986statistics}; to estimate the causal effect, we
need assumptions about the data generating process.

Assumptions are important to all probabilistic modeling, but
especially for making causal inferences. In such inferences, we make
strong assumptions about how treatments are assigned to individuals
and each individual's distribution of potential outcomes.
Given assumptions about the data, there are
myriad methods for model-based causal inference~\citep{pearl2000causality,robins2000marginal,morgan2014counterfactuals,imbens2015causal}.

We focus on Bayesian methods.
Bayesian methods have a long history in causal
inference~\citep{rubin1974estimating,raudenbush1986hierarchical}.
In recent years, applied researchers are increasingly able
to fit complicated probabilistic models to capture
fine-grained phenomena---for example, high-dimensional regressions
with regularization to handle large numbers of
predictors~\citep{maathuis2009estimating,belloni2014inference}, or hierarchical models to
adjust for unmeasured covariates and capture heterogeneity in
treatment effects~\citep{hirano2000assessing,feller2014hierarchical}.

However, for applied researchers, Bayesian methods can been difficult
to use.
One reason for this is the lack of diagnostic tools to check such
complicated models.
For high-dimensional and massive data, prior information---beliefs that capture our assumptions
on the data generating process---can be
essential in order to draw efficient causal inferences~\citep{bottou2013counterfactual,peters2015causal,johansson2016learning}. This in turn necessitates ways to check the modeling assumptions.

We develop model criticism for Bayesian
causal inference.
We build on the idea of
\textit{\glspl{ppc}}~\citep{box1980sampling,rubin1984bayesianly,gelman1996posterior}
to adapt goodness-of-fit style calculations to causal modeling.
We decompose the problem, separately criticizing the
two components that make up a causal model: the model of treatment
assignments and the model of outcomes.

We emphasize that there are several causal assumptions that our method
cannot check.  First, we do not check unconfoundedness.  There are
many existing methods for this, such as robustness tests~\citep{angrist2004treatment,lu2014robustness,chen2015exogeneity} and estimation of
pseudo causal effects~\citep{rosenbaum1987role,heckman1989choosing}. The results of these methods do not directly test unconfoundedness---such tests are theoretically not
possible~\citep{pearl2000causality}---but can lend plausibility to the
assumption. Second, we do not check correlation between potential outcomes.
This is also untestable as no pair of
potential outcomes is observed for a single data point. In practice,
we recommend using existing methods to understand the plausibility of
these untestable assumptions, and then to apply our methods to check
the additional assumptions.

In \mysec{causalmodels},
we describe model-based causal inference with potential
outcomes.
In \mysec{validation}, we develop methods for checking the models and confirm
their properties on simulated data. In \mysec{experiments},
we show how the methods can be applied for real-world problems: an observational study
on the effect of pest management in urban apartments and an
educational experiment on the effect of television exposure
on children.

\subsection{Related Work}

There has been little work on Bayesian model
criticism for causal inference.  Model checking is a frequent activity
in the practice of propensity score analysis, typically for
matching~\citep{dehejia2005practical,austin2009balance}.  Standard
econometric texts \citep{greene2003econometric,wooldridge2010econometric}
discuss regression
diagnostics and specification error. Neither of them considers a
Bayesian treatment of diagnostics or how to check Bayesian methods
for causal inference.

Our method borrows from the rich literature on doubly robust
estimation~\citep{robins2001comments,van2003unified,bang2005doubly}
and inverse probability of treatment
weighting~\citep{rosenbaum1983central,heckman1998matching,dehejia2002propensity}.
Such methods are designed to mitigate selection bias, which arises from
phenomena such as uncontrolled nonresponse and attrition, for the
estimation of causal effects. In \mysub{outcome}, we consider how such techniques
can be applied for diagnosing model misfit.

Central to \glspl{ppc} is the idea of predictive assessment: a model is evaluated by its predictions on future data given past
information~\citep{dawid1984present}. This has come up in the context
of causal inference via targeted learning and the super learner~\citep{van2006targeted,van2007super}. In fact, targeted learning
can be thought of as maximizing the statistical power for a given
\gls{ppc} (which we describe in \mysec{validation}).

In the context of missing data analysis, imputation of missing data has been a
de facto standard for model checking (see, e.g.,
\citet{chaloner1991bayesian,gelman2005multiple,su2011multiple}).  This
approach can also be applied for checking causal models.
In \mysec{experiments}, we
discuss it in detail and compare it to our approach.

\section{Causal models}
\label{sec:causalmodels}

We describe causal models in terms of the \textit{potential outcomes
framework}~\citep{neyman1923on,rubin1974estimating,imbens2015causal}.
Let $\mby(0), \mby(1)$ be the set of potential outcomes
under a binary treatment $\mba\in\{0,1\}$; and let $y_i(a)$ be the
outcome when an individual $i$ is exposed to treatment $a$. We use
uppercase, e.g., $\mbY(0),\mbY(1)$, to denote random variables and
lowercase, e.g., $\mby(0),\mby(1)$, to denote their realizations. (See
Appendix for a table of our notation.)

\subsection{Definition of a causal model}

Let $x_i$ denote a set of covariates for an individual $i$. The potential
outcomes arise from an \textit{outcome model},
\begin{align}
  \theta &\sim p(\theta) \nonumber \\
  x_i &\sim p(x) \label{eq:outcome-model} \\
  (y_i(0), y_i(1)) &\sim p(y(0), y(1) \g \theta, x_i). \nonumber
\end{align}
The potential outcomes are exchangeable across individuals; they are conditionally
independent given the outcome parameters $\theta$ and covariates $x_i$.

Unfortunately we do not observe both potential outcomes for any
individual.  The outcome we observe is determined by the
\textit{assignment model} of the treatment indicators $a_i$. The
treatment indicator $a_i$ equals one if we observe $y_i(1)$ and equals
zero if we observe $y_i(0)$.

We consider the unconfounded assignment model.  Each assignment is
drawn conditional on the covariates $x_i$ and unknown assignment
parameters $\phi$,
\begin{align}
  \begin{split}
  \phi & \sim p(\phi) \\
  a_i & \sim p(a \g x_i, \phi). \label{eq:obs-model}
  \end{split}
\end{align}
This model assumes \emph{unconfoundedness}: $(Y(0), Y(1)) \indep A \g X$. In other words,
conditional on the covariates, the potential outcomes are independent
of the treatment assignment.
In missing data analysis, the assumption
is known as strong ignorability~\citep{little1987statistical}.

We combine the outcome and assignment model in a \textit{causal
  model}.  With an unconfounded assignment model, the causal model is
\begin{align}
  & p(\theta, \phi, \mbx, \mby(0), \mby(1), \mba) = \\
  &\hspace{2em} p(\theta) p(\phi) p(\mbx)
  p(\mby(0), \mby(1) \g \theta, \mbx)
  p(\mba \g \phi, \mbx).  \nonumber
  \label{eq:causal-model}
\end{align}
We observe $\mbx$, $\mba$, and $\mby(\mba)$; all the other variables are
latent.

For exposition, we make a few simplifications. Specifically, we
assume:
(1) the outcomes of an individual are independent
of the assignment of other individuals;
(2) the outcome and assignment model parameters are independent;
and
(3) the treatment assignment has binary support.
The approach we describe extends beyond these settings.

\subsection{Bayesian inference in a causal model}

Given observed data $\{\mbx, \mba, \mby(\mba)\}$, we would like to calculate
the posterior distribution of the assignment parameters $\phi$ and
outcome parameters $\theta$.  Let $\bar{\mba}$ denote the unobserved
counterfactual assignments, i.e., $\bar{\mba} = 1 - \mba$. We marginalize out the counterfactuals to
calculate the posterior,
\begin{align}
  \begin{split}
  p(\theta, \phi \g \mby(\mba), \mba, \mbx) &\propto
  p(\theta) p(\phi) p(\mba \g \phi, \mbx) \\
  & \times \int p(\mby(\mba),
  \mby(\ensuremath{\bar{\mba}}) \g \theta, \mbx) \d{\mby(\ensuremath{\bar{\mba}})}.
  \end{split}
\end{align}
Because of unconfoundedness,
the posterior factorizes,\hspace{-2em}
\begin{align}
  & p(\theta \g \mby(\mba), \mba, \mbx)
  \propto
  p(\theta) \int p(\mby(\mba),
  \mby(\ensuremath{\bar{\mba}}) \g \theta, \mbx) \d{\mby(\ensuremath{\bar{\mba}})},
  \nonumber \\
  & p(\phi \g \mby(\mba), \mba, \mbx)
  \propto
  p(\phi) p(\mba \g \phi, \mbx).\nonumber
\end{align}
In other words, the assignment mechanism plays no role when inferring
potential outcomes. Similarly, the potential outcomes play no role
when inferring the assignment mechanism.
Motivated by this consequence of ignorability~\citep{rubin1976inference,dawid1977likelihood}, we
devise a method that separately checks the fitted parameters $\theta$ and $\phi$.

\section{Model criticism for causal inference}
\label{sec:validation}
\glsresetall

Model criticism measures the degree to which a model falsely describes
the data~\citep{gelman2012philosophy}.  We can never validate whether
a model is true---no model will be true in practice---but we can try
to uncover where the model goes wrong.  Model criticism helps justify
the model as an approximation or point to good directions for revising
the model.

The central tool of model criticism is the \gls{ppc}. It quantifies
the degree to which data generated from the model deviate from the observed data~\citep{box1980sampling,rubin1984bayesianly,gelman1996posterior}.

The procedure is:
\begin{enumerate}[leftmargin=*]
\item Design a \textit{discrepancy function}, a statistic of the data
  and hidden variables. A ``targeted'' discrepancy summarizes a
  specific component of the data, such as a quantile.  An ``omnibus''
  discrepancy is an overall summary of the data, such as the $\chi^2$
  goodness of fit.

\item Form the \textit{realized discrepancy}.  It is the discrepancy
  evaluated at the observed data along with posterior samples of the
  hidden variables.

\item Form the \textit{reference distribution}.  It the distribution
  of the discrepancy applied to data sets from the
  posterior predictive distribution.  In contrast to
  the realized discrepancy where the observations are fixed, the
  reference distribution is evaluated on samples of both observations
  and hidden variables.

\item Finally, locate the realized discrepancy in the reference
  distribution, e.g., by making a plot or by calculating a tail
  probability.  If the realized discrepancy is unlikely, then the model
  poorly describes this function of the data and we revise the model.
  If it is reasonable, then this provides evidence that
  the model is justified.

\end{enumerate}
\glspl{ppc} are typically applied to validating non-causal models,
especially for exploratory and unsupervised
tasks~\citep{yano2001evaluating,royle2008hierarchical,Mimno2011,Mimno2015}.  Here we
extend this methodology to validating causal models.

\subsection{\Acrlongpl{ppc} for causal models}

We first consider the discrepancy function.
Define a \emph{causal discrepancy} to be a scalar function of the
form,
\begin{align}
  T((\mby(0), \mby(1)), \mba, \theta, \phi).
\label{eq:causal-discrepancy}
\end{align}
It is a function of all variables in the causal model of
\myeqp{causal-model}: the potential outcomes $\mby(0),\mby(1)$, the
treatment assignment $\mba$, the outcome model parameters $\theta$,
and the assignment model parameters $\phi$.  Depending on the check,
it is a function of a subset of these variables.

In the original formulation of a \gls{ppc}, the discrepancy was a
function solely of observed data~\citep{rubin1984bayesianly}.  Later
work extended the discrepancy to also depend on latent
parameters~\citep{meng1994posterior,gelman1996posterior}.  The causal
discrepancy of \myeqp{causal-discrepancy} depends on observed
outcomes and assignments, latent parameters, and unobserved outcomes,
i.e., the counterfactual outcomes. Discrepancies of this form were
studied in the context of missing data by \citet{gelman2005multiple}.

We will use causal discrepancies to check each piece of the causal
model.  To complete the definition of a causal check, we must define
the \emph{reference distribution} and the \emph{realized discrepancy}.
In a \gls{ppc} the reference distribution is the posterior
predictive.  This is the distribution that the data would have come
from if the model were true.

Let $\mby(0)^{\rm rep},\mby(1)^{\rm rep},\mba^{\rm rep}$ denote
replicated data from the posterior predictive distribution.
Define the observed dataset $\cD^{\rm obs} = \{\mbx, \mby(\mba), \mba\}$.
We
replicate the assignments and outcomes conditioned on $\cD^{\rm obs}$,
\begin{align}
  \begin{split}
    \phi^{\rm rep} &\sim p(\phi\g\cD^{\rm obs}) \\
    \mba^{\rm rep} &\sim p(\mba \g\mbx,\phi^{\rm rep}) \\
    \theta^{\rm rep} &\sim p(\theta\g\cD^{\rm obs}) \\
    \mby(0)^{\rm rep},\mby(1)^{\rm rep} &\sim p(\mby(0),\mby(1) \g
    \mbx,\theta^{\rm rep}).
    \label{eq:reference-distribution}
  \end{split}
\end{align}
This defines the reference distribution of the causal discrepancy of
\myeqp{causal-discrepancy}.

A causal check compares the reference discrepancy to the realized
discrepancy.  The realized discrepancy is evaluated on observed data.
When $T(\cdot)$ depends on latent variables---either assignment
parameters, outcome parameters, or alternative outcomes---we replicate
them from the reference distribution.  (In that case the realization
of the discrepancy is itself a distribution.)  Following
\citet{gelman1996posterior,gelman2005multiple}, the observed data are
always held fixed at their observed values; only latent variables are
replicated.  This is in contrast to the reference distribution, which
samples all of the variables. We note that in causal inference, we cannot use this
strategy with the counterfactual outcomes; we will discuss this nuance
below in \mysub{outcome}.

We described the causal discrepancy, its reference distribution, and
its realization.  We now show how to use these ingredients to
criticize causal models. We separate criticism into two components:
criticizing the assignment model and criticizing the outcome model.

\subsection{Criticizing the assignment model}
\label{sub:assignment}

The gold standard for validating causal models is a held-out
experiment, where we have access to the assignment mechanism when
validating against held-out outcomes~\citep{rubin2008for}.  In
observational studies, however, the assignment mechanism is unknown;
we must model it with the goal of capturing the true distribution
of the assignments. To check this aspect of the model, we apply a standard
\gls{ppc}. Under the assumption of unconfoundedness, we can check the
assignment model with discrepancies $T(\mba, \phi)$ that are
functions of the assignment parameters $\phi$ and assignments $\mba$.

\myalg{assignment} describes the procedure.  It isolates the
components of the model and data relevant to the assignment mechanism.
First we calculate the realized discrepancy
$T(\mba, \phi^\textrm{\textrm{rep}})$; then we compare against the
reference distribution $T(\mba^{\rep}, \phi^{\rep})$.  The reference
distribution is simply the posterior predictive (\myeqp{reference-distribution}).\footnote{The realized
  discrepancy can also be evaluated on held-out assignments to avoid
  criticizing the model with the same observational data that is used
  to train it~\citep{bayarri2007bayesian}. This is the approach we use
  in our study.}

\begin{algorithm}[t]
  \caption{Criticism of the assignment model}
  \SetAlgoLined
  \DontPrintSemicolon
  \KwIn{Assignment model $p(\phi\g\cD^{\rm obs})p(\mba^{\rm rep}\g\mbx,\phi)$,
        Discrepancy $T(\mba, \phi)$.}
  \textbf{Output}: Reference distribution $p(T)$ and \\
                   \hspace{3.5em} Realized discrepancy $T^{\rm obs}$.\;
  \For{\textnormal{$s=1,\ldots,S$ replications}}{
    Draw assignment parameters $\phi^{s}\sim p(\phi\g\cD^{\rm obs})$.\;
    \vspace{0.4ex}
    Draw assignments $\mba^{{\rm rep},s}\sim p(\mba^{\rm rep}\g \mbx,\phi^{s})$.\;
    \vspace{0.4ex}
    Calculate discrepancy $T^{{\rm rep}, s} = T(\mba^{{\rm rep},s},\phi^{s})$.\;
    \vspace{0.4ex}
    Calculate discrepancy $T^{{\rm obs}, s} = T(\mba, \phi^{s}).$
  }
  Form reference distribution $p(T)$ from replications $\{T^{{\rm rep}, s}\}$.\;
  Form realized discrepancy $T^{\rm obs}$ from replications $\{T^{{\rm obs}, s}\}$.\;
  \label{alg:assignment}
\end{algorithm}

\begin{figure*}[t]
	\centering
	\begin{subfigure}[t]{400pt}
		\centering
		\includegraphics[width=0.26\textwidth]{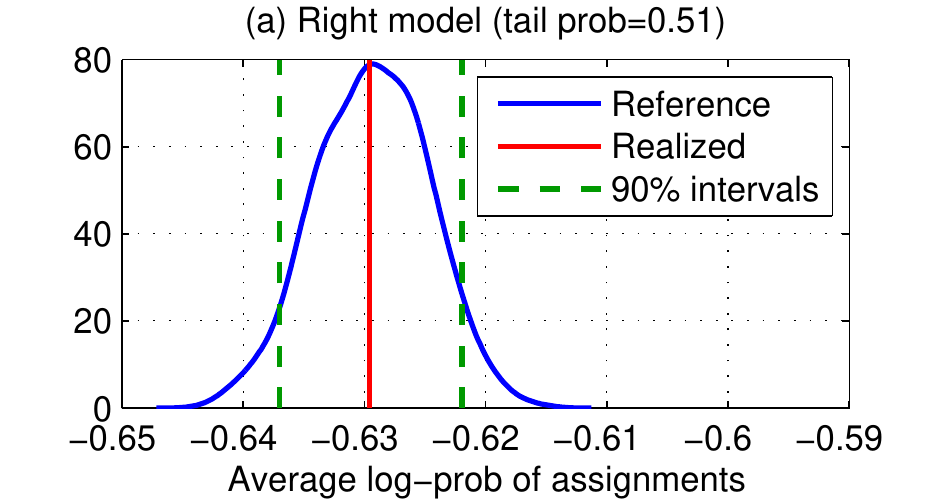} \hspace*{-14pt}
		\includegraphics[width=0.26\textwidth]{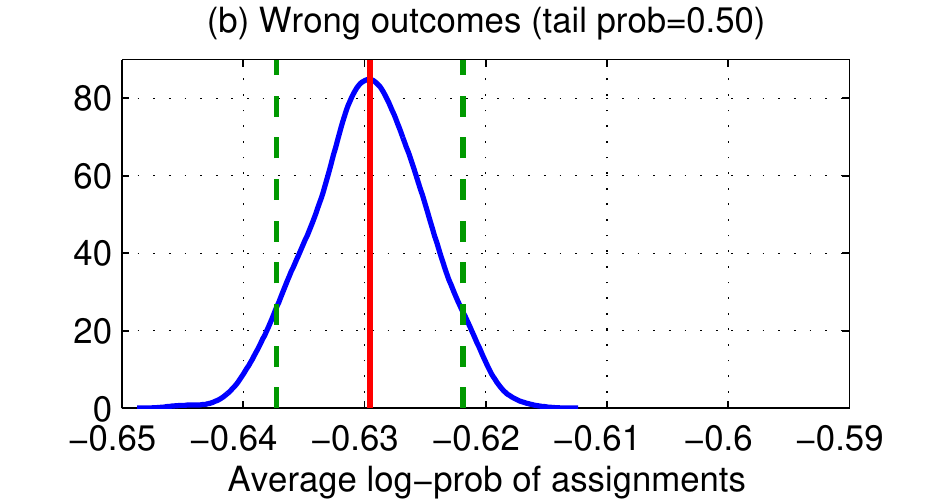} \hspace*{-14pt}
		\includegraphics[width=0.26\textwidth]{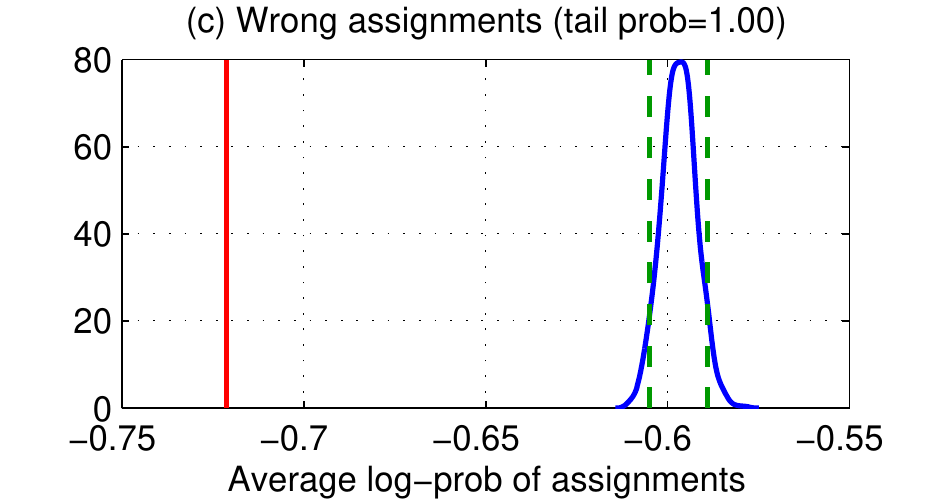} \hspace*{-14pt}
		\caption{\label{fig:sf_logp}Results of the assignment test. Model (c), which has a wrong assignment mechanism, fails the test. The plots for the fiction scenario (not shown) are similar to these ones, as the assignments are assumed independent of the outcomes.}
	\end{subfigure}
	\begin{subfigure}[t]{400pt}
		\centering
		\includegraphics[width=0.26\textwidth]{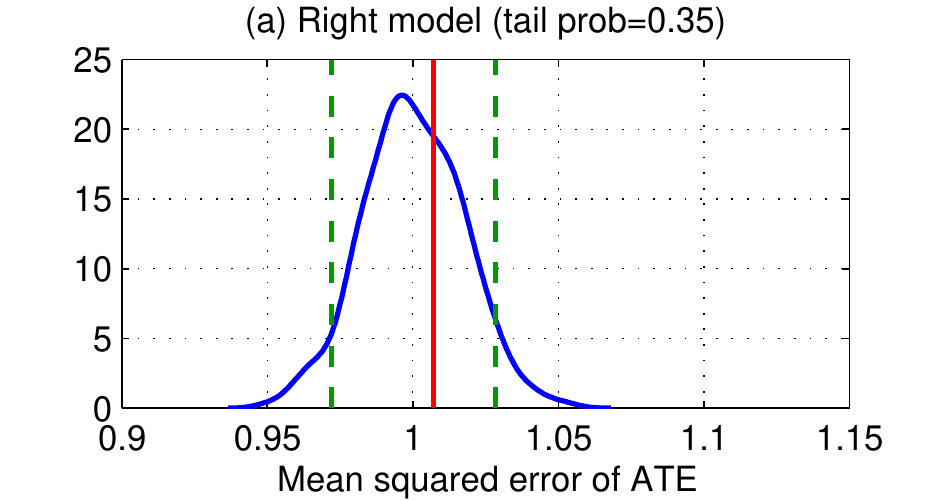}\hspace*{-14pt}
		\includegraphics[width=0.26\textwidth]{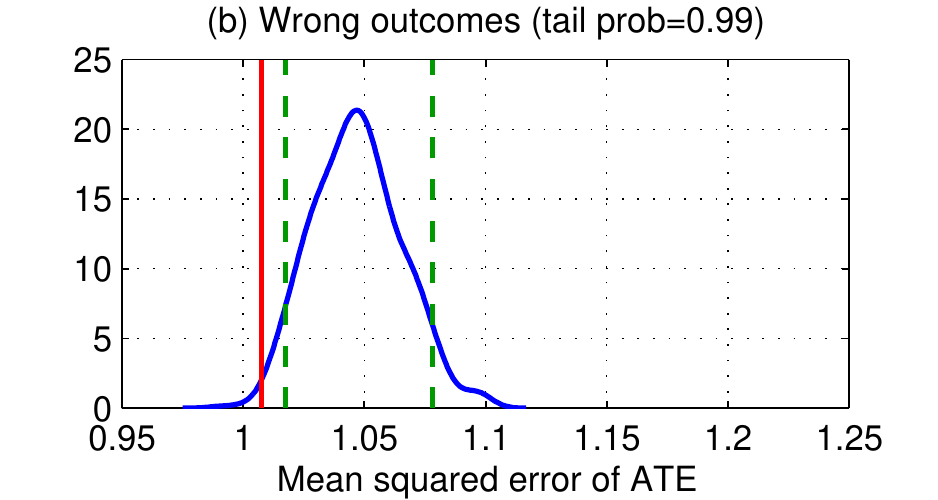}\hspace*{-14pt}
		\includegraphics[width=0.26\textwidth]{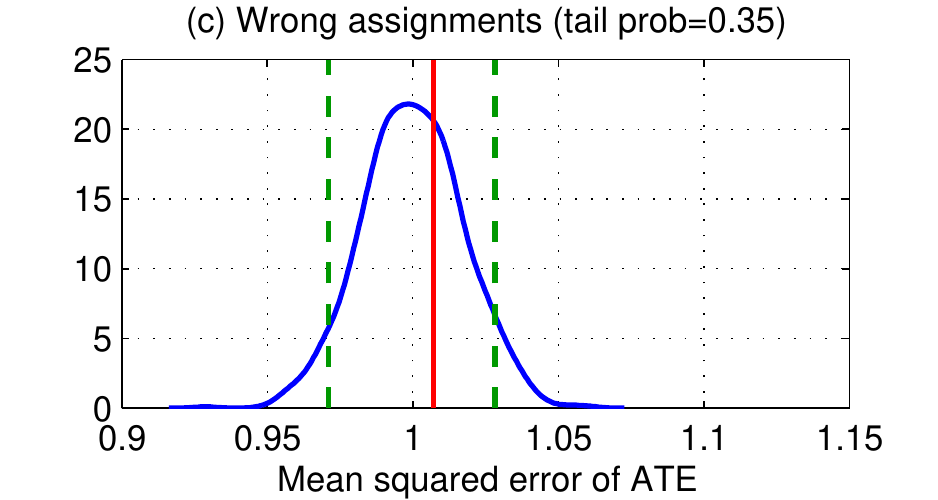}\hspace*{-14pt}
		\caption{\label{fig:sf_mse}Results of the outcome test. The test fails for the model in which the outcome model is misspecified.}
	\end{subfigure}
	\vspace*{-8pt}
	\caption{Results of the tests for the science-fiction scenario, in which we have access to both counterfactual outcomes.\label{fig:synthetic_sf}}
	\vspace*{-5pt}
\end{figure*}

\parhead{Example.}  Consider the average (marginal) log-likelihood of assignment,
\begin{align}
\begin{split}
T(\mba)
&= \frac{1}{n} \sum_{i=1}^n \log \int p(a_i\g x_i,\phi)p(\phi\g \cD^{\rm obs})d\phi.
\end{split}
\label{eq:log-prob}
\end{align}
The reference distribution is the log likelihood evaluated on
replications $\mba^{\rm rep}$. The realized discrepancy
is the log likelihood evaluated on the observed set of
assignments $\mba$.
\myfig{sf_logp} show examples of this check.  In the third panel we
consider a misspecified assignment model. The
log-likelihood (\red{red}) is far from its reference distribution.

\subsection{Criticizing the outcome model}
\label{sub:outcome}

The second component of a causal model is the outcome model,
\begin{align}
  p(\theta) p(\mby(0),\mby(1),\theta\g \mbx).
\end{align}
The outcome model represents the causal phenomenon, that is, the
distribution of the outcome $\mby(\mba)$ caused by the assignment $\mba$.  The outcome model is inherently difficult to infer and
criticize from data.  It involves inferences about a distribution of
counterfactuals, but with data only available from one counterfactual
world.

\parhead{Outcome discrepancies.}  We check an outcome model with an
outcome discrepancy, $T(\mby(0), \mby(1), \theta)$, a function of the
potential outcomes and the parameters that govern the outcome model.
One simple example is the average log-likelihood of the potential outcomes,
\begin{align}
  \frac{1}{n}\sum_{i=1}^{n}\left( \log p(y_i(0) \g \theta, x_i) + \log p(y_i(1) \g
  \theta, x_i) \right)
  \label{eq:log-prob-outcome}
\end{align}
Another example (used, e.g., by \citet{athey2015machine}) is an adjusted
mean squared error of the average treatment effect,
\begin{align}
  \begin{split}
  \frac{1}{n}\sum_{i=1}^n \Big( &(y_i(1) - y_i(0) - \E{Y_i(1) - Y_i(0) \g \theta,
  x_i})^2 \\
  & - (y_i(1) - y_i(0))^2 \Big).
  \label{eq:ate}
  \end{split}
\end{align}

\parhead{Propensity weighting for the realized discrepancy.}
Unlike the assignment checks, the outcome discrepancy is a
function of counterfactual outcomes $\mby(\bar{\mba})$, which we
do not observe. One approach is to use posterior samples of $\theta$ to impute these
counterfactuals~\citep{gelman2005multiple}.  However, this is not
appropriate when making a causal check. We discuss this nuance after
deriving an alternative approach.

As an alternative, we use a strategy based on propensity
scores~\citep{rosenbaum1983central} and doubly robust estimation~\citep{bang2005doubly}.
Consider discrepancies that are
sums of functions of the individual outcomes and outcome parameter,
\begin{align}
  & T(\mby(0), \mby(1), \theta) \label{eq:sum-discrepancy} \\
  & = \sum_{i=1}^{n} \left( f^0(y_i(0), \theta) +
  f^1(y_i(1), \theta) \right). \nonumber
\end{align}
For example, in \myeqp{log-prob-outcome},
\begin{equation*}
f^1(y_i(1), \theta) = \log p(y_i(1) \g \theta).
\end{equation*}
(To minimize notation, we assume that the functions are identical across data
points. They can also depend on the index $i$.)
For now, we focus on the treatment term $f^1(y_i(1), \theta)$; the
control term is analogous.  Consider an intervention $\delta_1(a)$
that always assigns $a$ to the treatment, that is, it places
probability one on $a = 1$ and probability zero on $a = 0$. The treatment term
can be rewritten as an expectation under the distribution
$a_i \sim \delta_1(a)$,
\begin{align}
  f^1(y(1), \theta) = \EE{\delta_1}{\ind[A_i = 1] f^1(y_i(1), \theta)}.
  \label{eq:delta-form}
\end{align}
Of course, we did not observe $a_i$ from this delta distribution.  So
we approximate the expectation with an importance weight,
\begin{align}
  f^1(y(1), \theta) \approx \frac{\delta_{1}(a_i)}{\pi_i(a_i)}
  f^1(y_i(1), \theta).
  \label{eq:ipw-term}
\end{align}
The denominator $\pi_i(a_i)$ is the (marginal) probability of the assignment under
the causal model,
\begin{align}
  \pi_i(a_i) = \int p(a_i \g \phi) p(\phi \g \mba, \mbx) d\phi.
  \label{eq:assignment-predictive}
\end{align}

\myeqp{ipw-term} only depends on observed data.  It equals zero when
the function depends on a counterfactual outcome that we do not
observe, i.e., when $a_i = 0$.  It is non-zero when $a_i = 1$ and we
observe $y_i(1)$. It is always a valid approximation but note it
assumes that the assignment probability is correct, i.e., that the
assignment model is accurate.

We apply this approximation to each term in \myeqp{sum-discrepancy}.
This gives an estimate of the realized discrepancy that only
depends on observed data,
\begin{align}
  \label{eq:ipw-discrepancy}
  & T(\mby(0), \mby(1), \theta) \approx \\
  & \hspace{2em} \sum_{i=1}^{n} \left ( \frac{\delta_1(a_i)}{\pi_i(a_i)} f^0_i(y_i(0), \theta) +
  \frac{\delta_0(a_i)}{\pi_i(a_i)} f^1(y_i(1), \theta) \right).\nonumber
\end{align}
\myalg{outcome} summarizes the procedure.
This strategy---replacing each function with its inverse probability weighted
realization---is appropriate beyond sums of functions of the outcomes.
It also works for polynomials of functions and sums of such
polynomials.  This is the setting of the discrepancy in \myeqp{ate}.

\begin{algorithm}[t]
  \caption{Criticism of the outcome model}
  \SetAlgoLined
  \DontPrintSemicolon
  \KwIn{Causal model
        $p(\mbtheta\g\cD^{\rm obs})
         p(\mbphi\g\cD^{\rm obs})
         p(\mby(0),$\\$\mby(1)\g\mbx,\mbtheta)
         p(\mba\g\mbx,\mbphi)$,\\
        \hspace{2.8em}
        Discrepancy $T((\mby(0), \mby(1)),\mbtheta)$.}
  \textbf{Output}: Reference distribution $p(T)$ and \\
                   \hspace{3.5em} Realized discrepancy $T^{\rm obs}$.\;
  \For{\textnormal{$s=1,\ldots,S$ replications}}{
    Draw outcome parameters $\mbtheta^{s}\sim p(\mbtheta\g\cD^{\rm obs})$.\;
    \vspace{0.4ex}
    Draw outcomes $\mby(0)^{{\rm rep},s},\mby(1)^{{\rm rep},s}\sim p(\mby(0)^{\rm
    rep},\mby(1)^{\rm rep}\g\mbx,\mbtheta^{s})$.\;
    \vspace{0.2ex}
    Calculate discrepancy
    $T^{{\rm rep}, s} = T((y(0)^{\rm rep},y(1)^{\rm rep}),\mbtheta^{s})$.\;
    \vspace{0.2ex}
    Calculate discrepancy
    $T^{{\rm obs}, s}$ using \myeqp{ipw-discrepancy}.
  }
  Form reference distribution $p(T)$ from replications $\{T^{{\rm rep}, s}\}$.\;
  Form realized discrepancy $T^{\rm obs}$ from replications $\{T^{{\rm obs}, s}\}$.\;
  \label{alg:outcome}
\end{algorithm}

\parhead{Analyzing bias: outcome checks with imputation versus importance reweighting.}
We now explain why imputation is not
appropriate when making a causal check.
First, by definition, imputed
values are well described by the posterior sample of $\theta$.  Thus,
checks according to the realized discrepancy only deviate via the
observed data, ignoring the process by which the values need to be
imputed.  Second, all such checks are inherently conditional on a
given set of treatment assignments $\mba$. Thus, they may not generalize
well for other treatment assignments.

We now show formally that imputation of missing data
results in biased estimates of the
counterfactual terms.
Consider a discrepancy
as in \myeqp{sum-discrepancy}, and
focus on a single term, $f(y_i(1), \theta)$.
Imputation replaces unobserved counterfactuals
$y(\bar{a}_i)$ with random outcomes drawn from $\theta$.  Denote their
mean $\mu_i(a) = \E{f(Y(1), \theta) \g \theta, x_i}$.

The difference between
the approximation by \citet{gelman2005multiple} and the truth is
\begin{equation*}
\ind[a_i = 1] f(y_i(1), \theta) + \ind[a_i = 0] \mu_i(1) - f(y_i(1), \theta).
\end{equation*}
In other words, the approximation applies $f$ to the observed $y_i(1)$ if $a_i=1$, and it imputes the value if $a_i=0$.

Let $\mu^*_i(1)$ be the true expectation under $Y(1)$.
Define bias to be the expectation of the difference
under the true causal model.
After some algebra, it is
\begin{align}
  (\pi_i(1) - 1) (\mu^*_i(1) - \mu_i(1)),
\end{align}
where as before, $\pi_i(1)$ is the probability of treatment $p(a_i = 1)$.

This bias is non-zero except when $\mu_i(1) = \mu^*_i(1)$, i.e., when
the outcome model is correct, or when the assignment mechanism is deterministic ($\pi_i = 1$). In the usual scenario with $\pi_i \neq 1$, we will incur
bias in the counterfactual term. Assuming the outcome model is
correct does not make sense when we are checking the outcome model.

In contrast, consider the importance-weighted estimate. The difference
between the approximation and the true value is
\begin{align}
  \frac{\ind[a_i = 1]}{\pi_i(1)} f(y_i(1), \theta) - f(y_i(1), \theta).
\end{align}
Under the assignment model, this has expectation zero.
The importance weighted estimate is unbiased.


\begin{figure*}
	\centering
	\begin{subfigure}[t]{300pt}
		\centering
		\includegraphics[width=100pt]{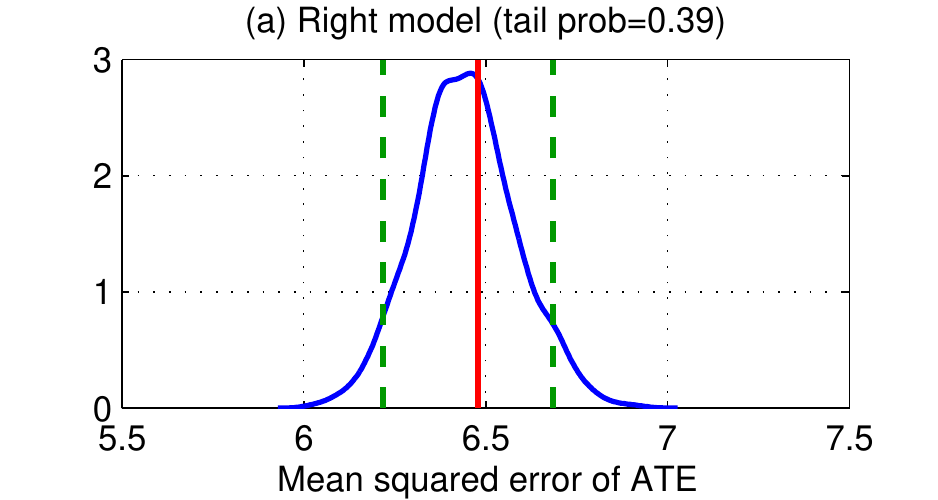}\hspace*{-14pt}
		\includegraphics[width=100pt]{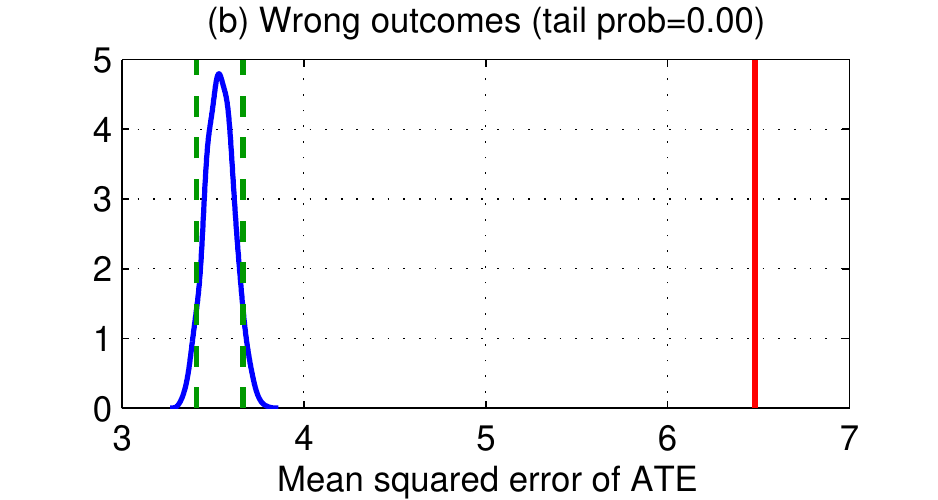}\hspace*{-14pt}
		\includegraphics[width=100pt]{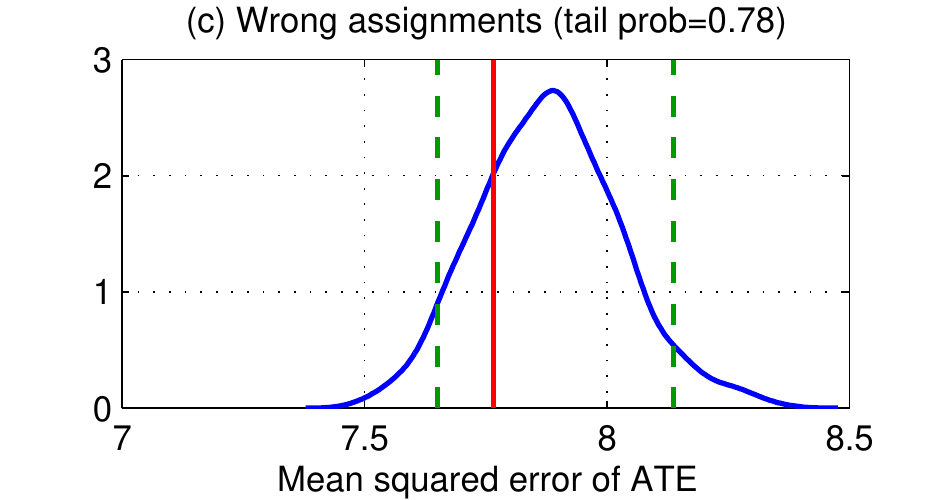}\hspace*{-14pt}
		\caption{\label{fig:f_mse}Our approach.}
	\end{subfigure}\hspace*{-10pt}
	\begin{subfigure}[t]{100pt}
		\centering
		\includegraphics[width=100pt]{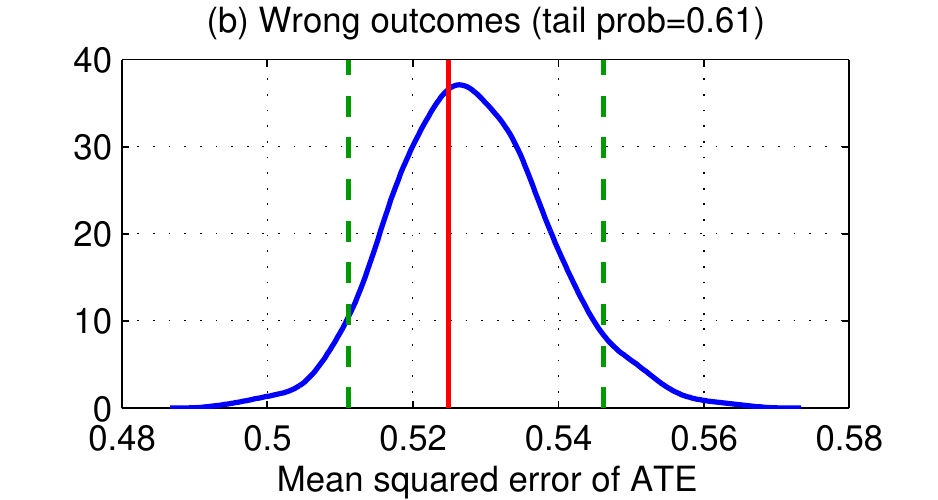}\hspace*{-14pt}
		\caption{\label{fig:f_mse_gelman}\citet{gelman2005multiple}.}
	\end{subfigure}
	\vspace*{-8pt}
	\caption{Results of the outcome tests for the fiction scenario, in which we only observe one counterfactual outcome. (a) The outcome test fails for the model in which the outcome model is misspecified. (b) If we impute the missing data following \citet{gelman2005multiple}, the test does not fail, although we are using the wrong outcome model.
	\label{fig:synthetic_f}}
	\vspace*{-5pt}
\end{figure*}




\section{Empirical Study}
\label{sec:experiments}

We use both synthetic and real data to illustrate how to validate
causal models. With synthetic data we compare our conclusions to the
true
data generating mechanism. With real data we demonstrate
our approach to criticize causal models in practice.  In all
our studies, we apply the discrepancies from
\mysub{assignment}; \mysub{outcome}, i.e., the marginal log-likelihood of the
assignments (\myeqp{log-prob}) and the mean squared error of the
\acrlong{ate} (\myeqp{ate}). Note we focus
here on the insights of the criticisms and not the insights of the
causal inferences.

\subsection{Synthetic data}

We showcase the results of different predictive checks using data
generated from a linear model (detailed below).
We first perform inference under the correct model
specification. Then we introduce misspecifications, either on the
outcome model or the assignment model.

We generate $10,000$ data points, each with a $10$-dimensional covariate
$x_i$, a binary treatment $a_i$, and a set of potential outcomes
$(y_i(0), y_i(1))$,
\begin{align*}
  & x_{i} \sim \textrm{Uniform}(x_i \g [0,1]^{10}), \\
  & a_i\g x_{i} \sim \textrm{Bernoulli}\left(a_i\g \textrm{logistic}(x_{i}^\top \phi)\right), \\
  & y_i(a) \g x_{i} \sim \mathcal{N}\left(y_i(a) \g [x_{i}, a]^\top \theta, \sigma^2 \right),
\end{align*}
We place a standard normal prior over the parameters $\phi$,
$\theta$, and a gamma prior with unit shape
and rate on $\sigma^2$.

We study two scenarios: (i) In the ``science-fiction'' scenario, we
have access simultaneously to $y_i(0)$ and $y_i(1)$. (This is not possible in the real world.) (ii) In the ``fiction''
scenario, we only have access to one counterfactual outcome,
$y_i=a_iy_i(1)+(1-a_i)y_i(0)$, and there are no hidden confounders. In each scenario, we check three causal
models: (a) the correct model as specified above; (b) a ``wrong
outcomes'' model, where we misspecify the distribution over
$y_i(a)$ by ignoring the first entry of $x_i$ (i.e., $\theta_1=0$);
(c) a
``wrong assignment'' model, where we misspecify the distribution over $a_i$ by
setting the probability that $a_i=1$ to $(0.7+0.3\times\textrm{logistic}(x_{i}^\top \phi))$.

We approximate the posterior with Markov chain Monte Carlo in
Stan~\citep{Carpenter2016}, obtaining $1000$ samples.
In the fiction scenario, we weigh the observations to form the outcome
discrepancy (\mysec{validation}). In the science-fiction
scenario, we do not weigh the observations because we see both
$y(0)$ and $y(1)$.

\myfig{sf_logp} illustrates criticism of the assignment
model (\myeqp{log-prob}) for the science-fiction scenario.
(The plots for the fiction scenario are similar.)
As expected, when we use the correct model
the realized discrepancy is approximately in the center of the
reference distribution (panel a); this indicates that the model is
correctly specified. We see the same for the misspecified outcome model
(panel b) because the assignment mechanism is still correct. However,
when the assignment model is wrong, the realized
discrepancy against the reference distribution (panel c) suggests
that the model is misspecified.

We now turn to criticizing the outcome model.
\myfig{sf_mse}; \myfig{f_mse} illustrate the test of the mean squared
error of the \gls{ate} (\myeqp{ate}) for the
science-fiction and fiction scenarios, respectively. As expected, the
test for the correct model does not suggest any issue (panel a). When
the outcome model is wrong, the test fails (panel b). This correctly
indicates that we should revise the model. In the science-fiction
scenario, the misspecification on the assignment model does not affect
the outcome model (panel c), and thus the test indicates correctness.

In the fiction scenario, our test for the outcome model indicates correctness.
In general, however,
the outcome test may
fail if the assignment model is misspecified because
it affects the inverse propensity weighting.

Finally,
\myfig{f_mse_gelman} shows the outcome test when we impute the missing
data following \citet{gelman2005multiple}. The bias leads the
test to pass (see \mysec{validation}), even though we use the misspecified
outcome model.

\subsection{Observational study: Cockroaches}
We analyze a real observational study of the
effect of pest management on cockroach levels in urban
apartments~\citep{Gelman2006}. Each apartment $i$ was set up with
traps for several days. The response is the number of
cockroaches trapped during that period; the treatment corresponds
to having applied the pesticide. We expect the pesticide to reduce
the number of cockroaches.

Let $t_{i}$ be the number of trap days and $y_i$ the number of
cockroaches for each apartment. We use two additional covariates: the
pre-treatment roach level $r_{i}$ and an indicator $s_i$ of whether
the apartment is a ``senior'' building, restricted to the elderly.  We
model the data as Poisson,
\begin{align*}
y_i\g a_i,x_i \sim \textrm{Poisson}\left(y_i\g \mu_i\right),
\end{align*}
where $\mu_i=t_i\exp\left\{\theta_0+\theta_1 s_i + \theta_2 r_i +
\theta_3 a_i\right\}$, $a_i$ is the treatment indicator, $\theta$ is
the outcome model parameter,
and $x_i=\{t_i,s_i,r_i\}$. We posit a logistic assignment model,
\begin{align*}
  a_i\g x_i \sim \textrm{Bernoulli}\left(a_i\g \textrm{logistic}([1,s_i,r_i]^\top \phi)\right).
\end{align*}
We
place standard normal priors over the parameters and draw $1 000$
posterior samples. 

\begin{figure*}[t]
	\centering
	\begin{subfigure}[t]{100pt}
		\centering
		\includegraphics[width=102pt]{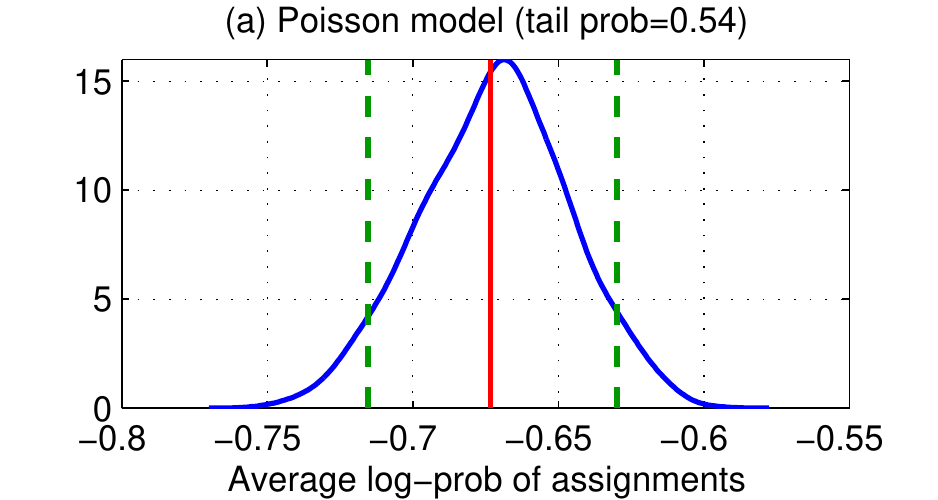}
		\caption{\label{fig:roaches_logp}Assignment test.}
	\end{subfigure}\hspace*{-10pt}
	\begin{subfigure}[t]{300pt}
		\centering
		\includegraphics[width=103pt]{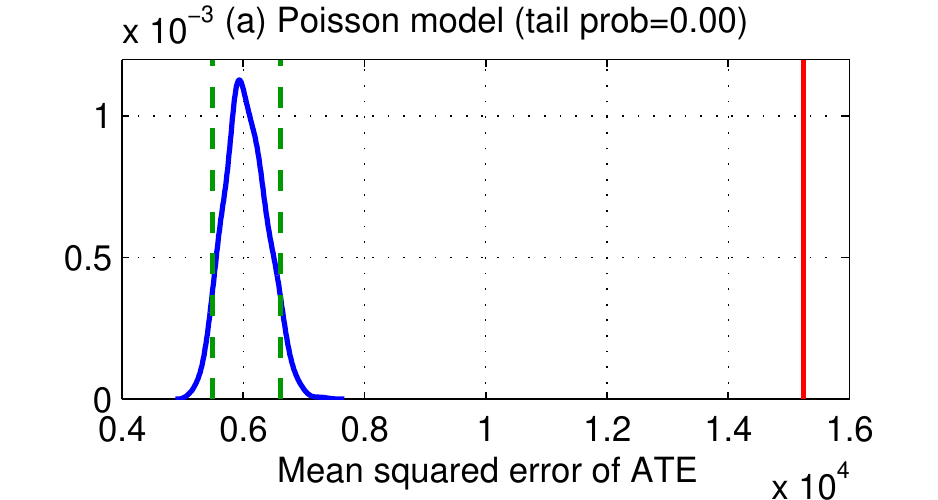}\hspace*{-15pt}
		\includegraphics[width=103pt]{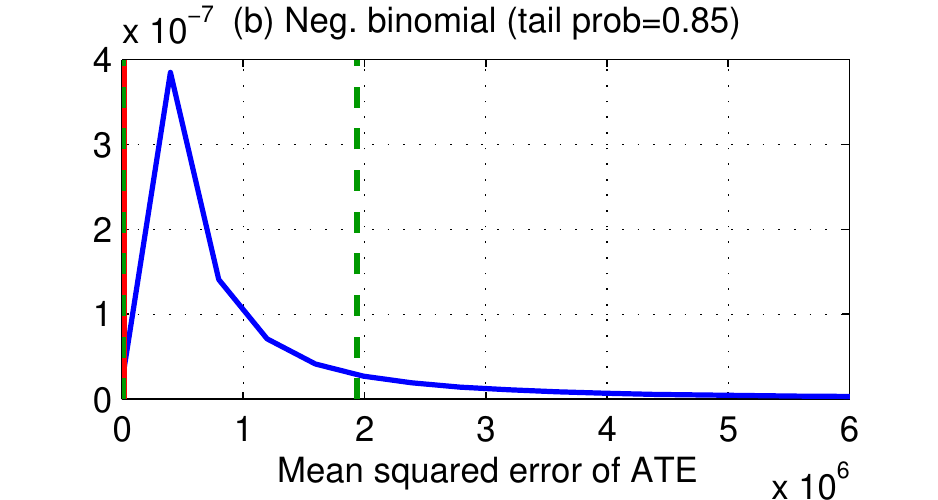}\hspace*{-15pt}
		\includegraphics[width=103pt]{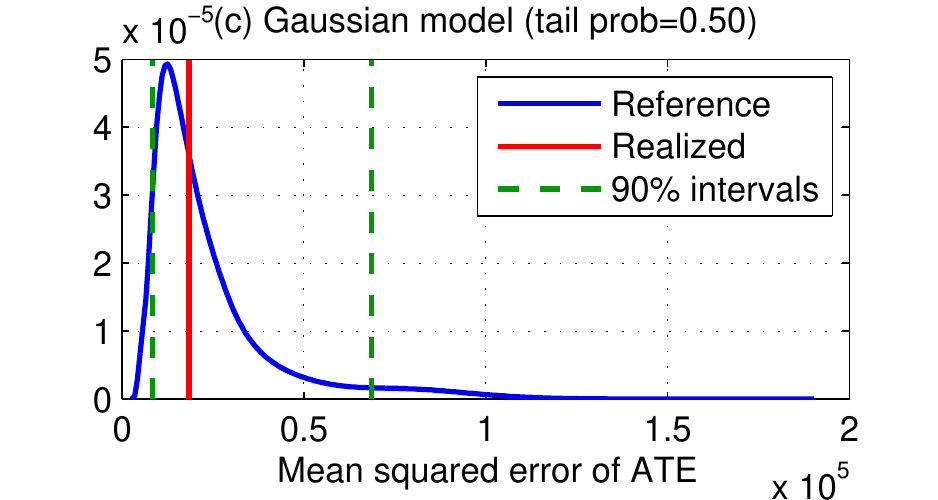}
		\caption{\label{fig:roaches_mse}Outcome test.}
	\end{subfigure}
	\vspace*{-3pt}
	\caption{\label{fig:roaches}Results for the cockroaches infestation study. The assignment test does not suggest any assignment model flaw. The outcome tests suggest that the variance is a linear function of the mean.}
	\vspace*{-8pt}
\end{figure*}

We first evaluate the assignment model with the average log-likelihood
of the assignments. Figure~\ref{fig:roaches_logp} illustrates the
test. The realized discrepancy is plausible.

Next we evaluate the
outcome model, again with the mean squared error of the
\gls{ate}. Figure~\ref{fig:roaches_mse} (left) illustrates the outcome
test for the Poisson model. The test fails: the model lacks the
overdispersion needed to capture the high variance of the
data~\citep{Gelman2006}. This is typical with the Poisson because its
variance is equal to its mean.

We propose two alternative models. Model (b) replaces the Poisson
likelihood with a negative binomial distribution.
It has the same mean $\mu_i$ as the Poisson but its variance increases
quadratically. The variance is $\mu_i+\mu_i^2/\theta_4$, where
$\theta_4$ is a dispersion parameter. We place a gamma prior with unit
shape and rate over $\theta_4$.  Model (c) has similar considerations,
but the variance is now a linear function of the mean
$\theta_4 \mu_i$.\footnote{This can be achieved with a quasi-Poisson
  regression~\citep{Gelman2006}, but this is not a proper
  probabilistic model. Rather, we use a heteroscedastic Gaussian
  distribution with the same mean and variance.}
\myfig{roaches_mse} (center and right) illustrates the causal tests
for models (b) and (c). These results suggest that model (c) is the
most plausible.

\begin{figure*}[tbh]
	\centering
	\includegraphics[width=0.3\textwidth]{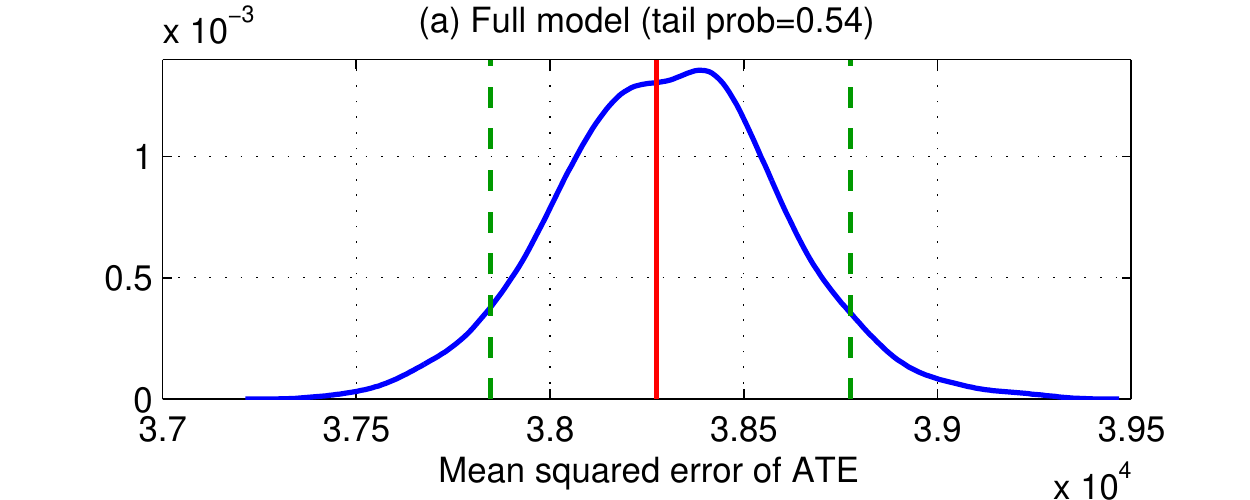}\hspace*{-25pt}
	\includegraphics[width=0.3\textwidth]{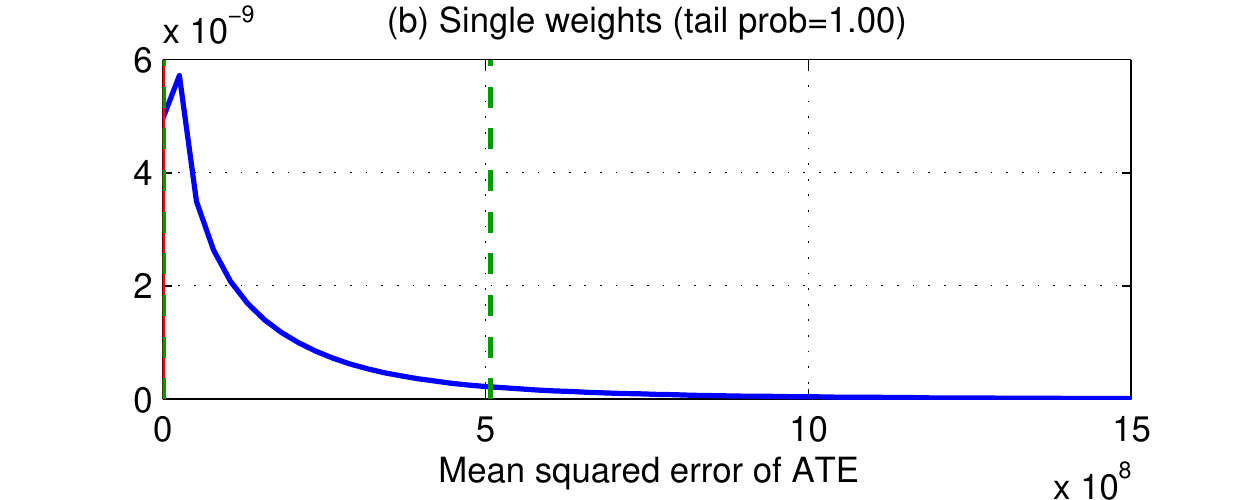}\hspace*{-25pt}
	\includegraphics[width=0.3\textwidth]{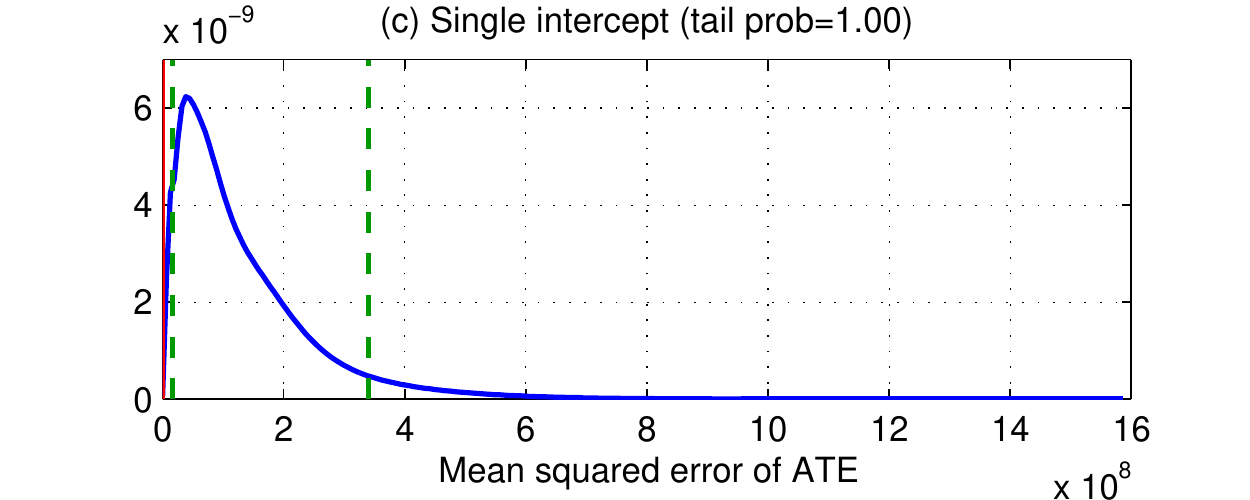}
	\caption{\label{fig:electric_mse}Results of the outcome test for the television show study. The models with a single weight for all grades (b) or a single intercept for all pairs (c) are too simple and fail the test.}
\end{figure*}

\subsection{Randomized experiment: The Electric Company television show}

We now consider an educational experiment performed around 1970 on a
set of elementary school classes. The treatment in this experiment was
exposure to a new educational television show called The Electric
Company. In each of four grades, the classes were completely
randomized into treated and control groups. At the end of the school
year, students in all the classes were given a reading test, and the
average test score within each class was recorded. The data are at the
classroom level.

Two classes from each grade were selected from each school. Let $y_{i1}$ and $y_{i2}$ be the scores of each class for the treatment and control groups, respectively. Let $p_{i1}$ and $p_{i2}$ be their pre-treatment scores at the beginning of the year. We also introduce the notation $g(i)$ to denote the grade (from 1 to 4) of the two classes from the $i$-th pair. We first use a regression model of the form
\begin{align*}
	& y_{i1} \sim \mathcal{N}(y_{i1}\g b_i + m_{g(i)} p_{i1} + \theta_{g(i)}, \sigma_{g(i)}^2), \\
	& y_{i2} \sim \mathcal{N}(y_{i2}\g b_i + m_{g(i)} p_{i2}, \sigma_{g(i)}^2),
\end{align*}
where the model parameters are: $b_i$, the intercept term that depends
on the specific pair $i$; $m_{1:4}$, the weight of $p_{i1}$ for each
grade; $\theta_{1:4}$, the treatment effect; and $\sigma^2_{1:4}$, the
variance for each grade. We place a Gaussian prior over the intercepts
$b_{i} \sim \mathcal{N}(\mu_{g(i)}, \tau_{g(i)}^2)$. We also place
Gaussian priors with zero mean and variance $10^4$ over $\mu_{g}$,
$\theta_g$, and $m_g$. We place gamma priors with shape $10$ and rate
$1$ over $\sigma_{g}$ and $\tau_{g}$. We refer to this model as model
(a). We also test two simplified models. Model (b) assumes that the
parameters $\theta$ and $m$ do not depend on the specific grade. Model (c) assumes instead that the intercept $b$ is shared for all pairs. Since we know that this is a completely randomized experiment by design, we do not posit any assignment model.

We plot in \myfig{electric_mse} the results of the outcome
test, which is based on the mean squared error of the average
treatment effect. Model (a), which is the most flexible, seems to
provide a sensible fit of the data. However, models (b) and (c) are
too simplistic. They clearly fail the test. If we had started from any
of these models in our analysis, the test would suggest the need to
revise them.

\section{Discussion}
\label{sec:discussion}

We have developed model criticism for Bayesian causal inference. This provides a rigorous foundation for diagnosing if a given probabilistic model is appropriate.

Here, we assume the typical setup in Bayesian causal
inference in which the posterior factorizes across outcome and assignment parameters.
However, this can lead to poor frequentist
properties~\citep{robins1997toward,robins2004optimal}.
To accommodate this, Bayesian-frequentist compromises have been proposed
which force dependency between the
outcome and assignment models
\citep{hoshino2008bayesian,mccandless2009bayesian,graham2016approximate}.
In future work, we will study causal discrepancies
which are of the general form of \myeqp{causal-discrepancy}, i.e.,
which depend simultaneously on assignments and potential
outcomes. Such discrepancies could also be applied to check causal
models with a non-ignorable assignment mechanism.

Finally, there has been a surge of interest in model-based causal
inference for high-dimensional, massive, and heterogenous data. In
such settings, one can capture more fine-grained phenomena, whether it
be with hierarchical
models~\citep{hirano2000assessing,feller2014hierarchical}, regularized
regressions~\citep{maathuis2009estimating,belloni2014inference},
structural equation
models~\citep{bottou2013counterfactual,peters2015causal}, or neural
networks~\citep{johansson2016learning}. This is an important regime
for checking causal models.

\section*{References}
\renewcommand{\bibsection}{}
\bibliographystyle{apalike}
\bibliography{bib}

\appendix

\section{Notation}

We provide a table describing the notation we use in this paper. See
\mytable{notation-data}; \mytable{notation-model}; \mytable{notation-criticism}.

\begin{figure}[t]
\begin{minipage}{\textwidth}
\centering
\begin{tabular}{ll}
\toprule
Symbol & Description\\
\midrule
$A_i$ & Treatment assignment of individual $i$ (random variable) \\
$Y_i(0),Y_i(1)$ & Potential outcomes of individual $i$ (random variable) \\
$\mbA=(A_1,\ldots,A_n)^\top$ & Set of treatment assignments (random variable) \\
$\mbY(0),\mbY(1)$ & Set of potential outcomes (random variable) \\
\hspace{0.5em}$=(Y_1(0),\ldots,Y_n(0))^\top,(Y_1(1),\ldots,Y_n(1))^\top$ \\
$a_i$ & Treatment assignment of individual $i$\\
$y_i(a_i)$ & Outcome of individual $i$ when assigned to treatment $a_i$
\\
$x_i$ & Observed covariates of individual $i$\\
$\mba$ & Set of treatment assignments \\
$\mby(0),\mby(1)$ & Set of potential outcomes \\
$\mby(\mba)=(y_1(a_1),\ldots,y_n(a_n))^\top$ & Set of outcomes when
assigned to set of treatments\\
$\mbx$ & Set of observed covariates \\
$\cD^{\rm obs} = \{a_i, y_i(a_i)\}$ & Observed data set \\
$\cD^{\rm do} = \{a_i^{\rm do}, y_i(a_i^{\rm do})\}$ & Hypothetical data set from an intervention \\
\bottomrule
\end{tabular}
\captionof{table}
{Notation for observational data.}
\label{table:notation-data}
\vspace{1ex}
\centering
\begin{tabular}{ll}
\toprule
Symbol & Description\\
\midrule
$\mbphi$ & Parameters of the assignment model \\
$p(a_i\g x_i,\mbphi)$ & Assignment likelihood for individual $i$\\
$p(\mba\g \mbx,\mbphi)
=\prod_{i=1}^n p(a_i\g x_i,\mbphi)$ & Assignment likelihood \\
$p(\mba\g \mbx,\mbphi)p(\mbphi)$ & Assignment model \\
$\mbtheta$ & Parameters of the outcome model \\
$p(y_i(0), y_i(1) \g x_i, \mbtheta)$ & Outcome likelihood for
individual $i$\\
$p(\mby(0), \mby(1) \g \mbx, \mbtheta)
=\prod_{i=1}^n p(y_i(0), y_i(1) \g x_i, \mbtheta)$ & Outcome likelihood \\
$p(\mby(0), \mby(1) \g \mbx, \mbtheta)p(\mbtheta)$ & Outcome model \\
$p(\mby(0),\mby(1)\g\mbx,\mbtheta)p(\mbtheta)p(\mba\g\mbx,\mbphi)p(\mbphi)$ & Causal model \\
$p(\mba^{\rm rep}\g\mbx,\mbphi)p(\mbphi\g\mba)$ & Assignment model (a posteriori) \\
$p(\mby(0)^{\rm rep},\mby(1)^{\rm rep}\g\mbx,\mbtheta)
         p(\mbtheta\g\mby)$ & Outcome model (a
         posteriori)\\
$p(\mby(0)^{\rm rep},\mby(1)^{\rm rep}\g\mbx,\mbtheta)
         p(\mbtheta\g\mby)
         p(\mba^{\rm rep}\g\mbx,\mbphi)
         p(\mbphi\g \mba)$ & Causal model (a posteriori) \\
\bottomrule
\end{tabular}
\captionof{table}
{Notation for causal models.}
\label{table:notation-model}
\vspace{1ex}
\centering
\begin{tabular}{ll}
\toprule
Symbol & Description\\
\midrule
$(\mby(0)^{\rm rep},\mby(1)^{\rm rep}),\mba^{\rm rep}$ & Replicated data set of
outcomes and assignments\\
$T((\mby(0), \mby(1)),\mba,\mbtheta,\mbphi)$ & Causal discrepancy
(over realizations) \\
$T^{\rm rep, s} = T((\mby(0)^{{\rm rep},s},\mby(1)^{{\rm rep},s}),\mba^{\rm
rep},\mbtheta^{s},\mbphi^{s})$ & Discrepancy over replication $s$ \\
$T^{\rm obs, s} = T((\mby(0),\mby(1)),\mba,\mbtheta^{s},\mbphi^{s})$ & Realized discrepancy over replication $s$\\
$p(T) = \{T^{\rm rep, s}\}$ & Reference distribution \\
$T^{\rm obs} = \{T^{\rm obs, s}\}$ & Realized discrepancy \\
\bottomrule
\end{tabular}
\captionof{table}
{Notation for model criticism.}
\label{table:notation-criticism}
\vspace{1ex}
\end{minipage}
\end{figure}

\end{document}